\makeatletter
\def\@biblabel#1{}
\makeatother

\documentclass[12pt]{iopart}

\newcommand{\aluminaM}{\ensuremath{\mbox{Al}_2\mbox{O}_3\mbox{:C,Mg}}}
\newcommand{\alumina}{\ensuremath{\mbox{Al}_2\mbox{O}_3}}
\newcommand{\degree}{\ensuremath{^\circ}}

\usepackage{graphicx} 
\begin{document}

\title{Ion track reconstruction in 3D using alumina-based fluorescent nuclear track detectors}

\author{M Niklas$^{1,2}$, J A Bartz$^{3,4}$, M S Akselrod$^4$, A Abollahi$^{2,5-8}$, O J\"{a}kel$^{1,2,6,7}$ and S Greilich$^{1,2}$}

\address{$^1$ German Cancer Research Center (DKFZ), Division of Medical Physics in Radiation Oncology, INF 280, 69120 Heidelberg, Germany, $^2$ German Cancer Consortium (DKTK), National Center for Radiation Research in Oncology, Heidelberg Institute of Radiation Oncology, INF450/400, Heidelberg Germany, $^3$ Oklahoma State University, Physics Department, Stillwater, OK 74078-3072, USA, $^4$ Landauer Inc., Stillwater  Crystal Growth Division, 723 1/2 Eastgate, Stillwater, OK 74074, USA, $^5$ Molecular \& Translational Radiation Oncology, Heidelberg Ion-Beam Therapy Center (HIT), University of Heidelberg Medical School and National Center for Tumor Diseases (NCT), German Cancer Research Center (DKFZ), 69120 Heidelberg, Germany, $^6$ Department of Radiation Oncology and Radiation Therapy, University Hospital Heidelberg, INF 400, 69120 Heideberg, Germany, $^7$ Heidelberg Ion-Beam Therapy Center (HIT), Im Neuenheimer Feld 450, 69120 Heidelberg, Germany, $^8$ Center of Cancer Systems Biology, Nasa Specialized Center Of Research (NSCOR), St. Elizabeth's Medical Center, Tufts University School of Medicine, Boston, MA, USA}

\ead{m.niklas@Dkfz-Heidelberg.de}

\begin{abstract}\\
Fluorescent nuclear track detectors (FNTDs) based on $\aluminaM$ single crystal combined with confocal microscopy provide 3D information on ion tracks with a resolution only limited by light diffraction. FNTDs are also ideal substrates to be coated with cells to engineer cell-fluorescent ion track hybrid detectors. This radiobiological tool enables a novel platform linking cell responses to physical dose deposition on a sub-cellular level in proton and heavy ion therapies. To achieve spatial correlation between single ion hits in the cell coating and its biological response the ion traversals have to be reconstructed in 3D using the depth information gained by the FNTD read-out. FNTDs were coated with a confluent human lung adenocarcinoma epithelial cell layer. Carbon ion irradiation of the hybrid detector was performed perpendicular and angular to the detector surface. In-situ imaging of the fluorescently labeled cell layer and the FNTD was performed in a sequential read-out. Making use of the trajectory information provided by the FNTD the accuracy of 3D track reconstruction of single particles traversing the hybrid detector was studied. The accuracy is strongly influenced by the irradiation angle and therefore by complexity of the FNTD signal. Perpendicular irradiation results in highest accuracy with error of smaller than 0.10$\degree$. The ability of FNTD technology to provide accurate 3D ion track reconstruction makes it a powerful tool for radiobiological investigations in clinical ion beams, either being used as a substrate to be coated with living tissue or being implanted in vivo.
\end{abstract}

\vspace{2pc}
\maketitle

\section{Introduction}
\begin{figure}[o]
	\centering
		\includegraphics{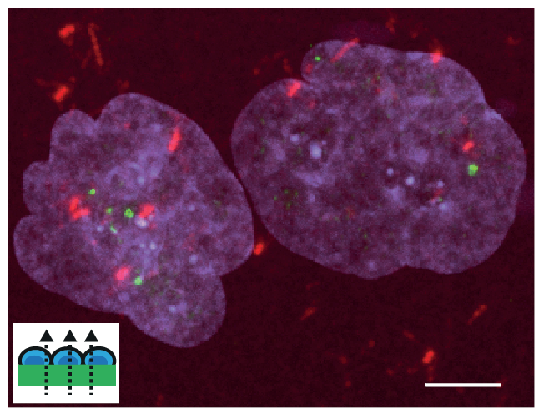}
	\caption{Spatial correlation between single carbon ion traversal and cell damage \cite{NiklasLumDetr2012}. FNTD was coated with A549 cells and irradiated perpendicularly to its surface (insert). Maximum intensity z projection between single carbon ion traversals (red spots) and subsequent DNA double strand breaks ($\gamma$-H2AX, immunofluorescent staining, green spots). Cell nuclei are labeled in blue (HOECHST 33342 staining). Scale bar, 5 $\mu$m.}
	\label{fig:correlation}  
\end{figure}

$\aluminaM$ based fluorescent nuclear track detectors (FNTDs) \cite{Akselrod2011} are ideal candidates for engineering cell-fluorescent ion track hybrid detectors \cite{NiklasRadOnc2012}. In this hybrid the FNTD is used as a substrate to be coated with a viable cell layer thus allowing for standard cellular imaging after proton and heavy ion irradiation. In addition the FNTD is used for single track detection. This radiobiological tool enables a novel platform linking physical energy deposition and biological response with a resolution only limited by diffraction (figure \ref{fig:correlation}). The major advantage of using FNTD as the central component is its capability of being read out by confocal laser scanning microscopy (CLSM) commonly used in life science \cite{Akselrod2011, Greilich2012}. In particular this allows for in-situ imaging of fluorescently labeled cell layer and the subjacent detector \cite{NiklasRadOnc2012}. The FNTD comprises superior spatial resolution compared to available CR-39 plastic nuclear track detectors and does not require chemical treatment \cite{Akselrod2006, Osinga2012}. FNTDs exhibit high detection efficiency (close to 100$\%$) \cite{Osinga2012}, a wide range of LET sensitivity \cite{Akselrod2011} and depth information \cite{Akselrod2006}.\\
In order to acquire accurate spatial correlation in 3D between individual ion hits in the cell layer and its subsequent response, the ion tracks have to be reconstructed in 3D using the depth information gained by the FNTD read-out \cite{Akselrod2006, Bartz2013}.\\
In this paper we analysed the FNTD signals after perpendicular and angular carbon ion irradiation of the hybrid detector. Making use of the trajectory information provided by the FNTD we studied the accuracy of 3D track reconstruction of particles traversing the hybrid detector. For this purpose single track spots - the ion's characteristic signature in the $\aluminaM$ - and their corresponding centers belonging to a single ion track were precisely located in the acquired FNTD image stack (a sequence of optical slices in z). Linear regression analysis was then applied to fit the ion track. Each track was extrapolated above the detector surface into a cell layer grown on top of the FNTD. Due to a refractive index mismatch in the optical path of the imaging of the hybrid detector correction of spherical aberrations \cite{Hell1993, Jacobsen1995} was applied.

\section{Materials and methods}

\subsection{$\aluminaM$ based FNTD}
FNTDs are made of alumina single crystals ($\alpha$-$\alumina$) doped with magnesium and carbon ions and exhibit high concentrations of $\mathrm{F_2^{2+}(2Mg)}$ aggregate defects (excitation at 435 nm and emission of fluorescence at 515 nm) \cite{Akselrod2011}. $\mathrm{F_2^{2+}(2Mg)}$ undergo radiochromic transformation. The resulting stable, transformed color centers, $\mathrm{F_2^{+}(2Mg)}$, absorb light in the band centered at 620 nm, prompting fast 750 nm fluorescence. The fluorescent intensity depends on local energy deposition of the ionizing radiation \cite{NiklasLumDetr2012}. Accordingly, FNTDs allow for particle track visualization by confocal microscopy and subsequent 3D particle track reconstruction \cite{Akselrod2006}. FNTDs are sensitive to ions with LET $>$ 0.5 keV/$\mu$m \cite{Akselrod2011}. The current limit of maximum countable track fluence is in the range of  $0.5 \cdot 10^8 \mathrm{cm^{-2}}$ \cite{Osinga2012} corresponding to clinical doses for carbon ion irradiation (i.e. 2.27 Gy in water for $^{12}$C, 90 MeV u$^{-1}$). FNTDs have one 4 mm by 8 mm surface polished to optical quality for read-out. The optical $c$-axis of the crystal is aligned parallel to the longer side of the detector (figure \ref{fig:setup}a). 

\subsection{Cell coating and irradiation setup}
The polished surface of sterilized FNTDs were coated with a confluent human lung adenocarcinoma epithelial (A549) cell layer with the protocol described in \cite{NiklasRadOnc2012} (plating density: 100 000 ml$^{-1}$, culture medium: Dulbecco's modified Eagle medium, Biochrom AG, Cat. No. FG 0415). A549 cells were obtained from Deutsche Sammlung von Mikroorganismen und Zellkulturen (DSMZ, Braunschweig, Germany). 15 minutes after the irradiation the cells remaining on the FNTD crystal were fixed with 4\% paraformaldehyde (PFA) in phosphate buffered saline (PBS) for 10 min at room temperature. Cell nuclei were stained with HOECHST 33342 fluorescent dye (Molecular Probes \textregistered, Cat. No. H1399, final concentration: 2 $\mu$g ml$^{-1}$) as described in \cite{NiklasRadOnc2012}.\\
Carbon ion irradiation of FNTD was performed using the therapy beam of the Heidelberg Ion Beam Therapy Center (HIT) at Heidelberg University Hospital. Irradiations were performed under the polar angle $\theta$ (angle between the direction of propagation of the ions $\vec{s}$ and the $k$-axis of $\aluminaM$) of $0 \degree \pm 5 \degree $ and $60 \degree \pm 5 \degree$ (figure \ref{fig:setup}a). The azimuth angle $\phi$ (angle between the optical c-axis and $\vec{e}_{x,y}$, the projection of the ion beam onto the exposed FNTD surface) only played a minor role for the setup. For perpendicular irradiation ($\theta= 0\degree$) the ion beam fluence was adjusted to $1.5 \cdot 10^6 \ \mathrm{cm^{-2}}$. For angular irradiation ($\theta= 60\degree$) the fluence at the FNTD surface corresponded to $1.3 \cdot 10^6 \ \mathrm{cm^{-2}}$. In both cases, a 12 x 12 cm$^2$ field was irradiated homogeneously using raster scanning with a pencil beam of 10.1 mm in diameter (full width at half maximum) and a distance of 2 mm between two raster spots. Approximately 60,000 particles were delivered in each spot. The Bragg peak was broadened in depth by using a 3 mm Ripple filter. The cell-coated FNTDs, mounted with agarose in a 24 multiwell plate filled with culture medium \cite{NiklasRadOnc2012}, were placed in the rising flank of the Bragg peak (initial carbon ion energy of 270.5 MeV u$^{-1}$, corresponding equivalent range in water $r_{H_{2}O}$= 13.70 cm). For the irradiation under $\theta= 0 \degree $ ($\theta= 60 \degree$) 11.70 cm (11.05 cm) of PMMA absorber with a corresponding $\mathrm{r_{H_{2}O}}$= 13.63 cm ($\mathrm{r_{H_{2}O}}$= 12.87 cm) was placed in front of the multiwell plate. For angular irradiation at $\theta= 60 \degree$ the multiwell plate was placed at an angle of $\gamma= 30\degree \pm 5 \degree$ ($\gamma= 90\degree - \theta$, defined for practical purpose) towards the incident ion beam. The different PMMA thicknesses result from the different thicknesses of the bottom of the multiwell plates (polystyrene, $\theta= 0 \degree$: $\mathrm{r_{H_{2}O}}$= 1.2 mm, $\theta= 60 \degree$: $\mathrm{r_{H_{2}O}}$= 2.5 mm). The air gap between the culture well and the PMMA was not considered in the total $\mathrm{r_{H_{2}O}}$. The amount of material in the beam from vacuum exit window to isocenter corresponds to a $\mathrm{r_{H_{2}O}}$ of 2.89 mm. 

\subsection{FNTD and cell layer read-out}
For the sequential read-out of the cell-coated FNTD we used the Zeiss LSM 710 ConfoCor 3 confocal microscopy equipped with a z-piezo stage, 63x/1.45 numerical aperture (NA) Oil DIC M27 objective, photomultiplier tubes (PMT) and avalanche photo diodes (APDs). We used the protocols as previously described in \cite{Greilich2012} and \cite{NiklasRadOnc2012}. For the read-out of the angular (perpendicular) irradiated FNTDs the excitation laser power of the 633 nm Helium-Neon laser was adjusted to 100$\%$ (100$\%$) transmission, pixel dwell time $\tau$ was set to 4.97 $\mu$s (2.80 $\mu$s) and the line-scanning repetition $R$ was limited to 4 (4). For the cell layer acquisition (HOECHST 33342) we used a 405 nm diode laser line (30 mW, 4$\%$ transmission) with $\tau$= 2.80 $\mu$s and $R$= 4. The microscope detector pinhole aperture was set to 1 Airy disk unit (AU). For angular (perpendicular) irradiated FNTDs a single imaging field comprised 1300 x 1300 pixel (1152 x 1152 pixel) with a pixel size of 0.104 x 0.104 $\mu \mathrm{m^2}$ (0.117 x 0.117 $\mu \mathrm{m^2}$). The acquired image stacks of the angular irradiated FNTDs covered an axial range of approximately 90 $\mu$m (measured from the detector surface). Concerning the perpendicular irradiated FNTD the image stacks covered an axial range of approximately 120 $\mu$m. In both cases the z-interval $\widehat{\Delta z}$ between two consecutive image planes was adjusted to 3 $\mu$m. For imaging, the cell-coated FNTD was placed in uncoated glass bottom culture dish (MatTek Corp., Part No. P35G-1.5-20C) with the cell-layer facing the glass bottom. The culture dish was filled with PBS. Anisotropic fluorescence properties of Al2O3:C,Mg crystals \cite{Sanyal2005,Greilich2012} were neglected. Zeiss Immersol$^\mathrm{TM}$518 F (n= 1.51 for $\lambda$= 643.8 nm at 23$\degree$C) was used as an immersion medium. The images acquired were stored with a bit depth of 16-bit in the LSM format.\\
The position of the polished FNTD surface was identified by using the HOECHST 33342 fluorescent signal from the nuclear staining. It disappears at the transition into the FNTD crystal. In addition, excitation by 405 nm causes a photoionization of the pristine $\mathrm{F_2^{2+}(2Mg)}$ aggregate defects located in close vicinity to the detector surface hence increasing the background in the HOECHST 33342 channel.

\subsection{Detection of the ion track centers}
For image segmentation the acquired 16-bit integer FNTD images were converted into floating point data (with pixel values in the interval $[0;1]$). A window limited to 30 x 30 pixels for perpendicular and 120 x 30 pixels for angular irradiation was used to define regions of interest (ROIs) within a plane of the FNTD image stack. Each ROI contained a single track spot. Thresholding in the ROIs was applied to identify all track spot centers in the acquired image stack belonging to a single particle trajectory. Due to secondary electron tracks sprouting from a track spot center each ROI can contain several local intensity maxima dividing it into several sub-areas. The largest area was isolated and considered to account for the ion track core (a halo of secondary electrons around the center of a track spot, figure \ref{fig:profile}a). The smaller areas mainly arise from dense energy depositions induced by strongly scattered secondary electrons having enough energy to leave the ion track core. We compared three different approaches to identify the center of a track spot:
\begin{enumerate}
	\item[(A)]{Detection of intensity-weighted centroids applying a global threshold (\itshape{iw centroid}\upshape): \\
	all pixels within the ROI with values greater than a manually defined global threshold of 0.4 were considered for the calculation of the intensity-weighted centroid.}
	\item[(B)]{Detection of intensity-weighted centroids applying a dynamic threshold (\itshape{rel thres}\upshape): \\
	the threshold was set to $2/3$ of the maximum pixel value within an ROI.}
	\item[(C)]{Detection of the absolute intensity maximum (\itshape{abs max}\upshape): \\
	the pixel with the maximum value within the ROI was considered as the track spot center.}
\end{enumerate} 
No background correction of the FNTD raw images was applied.

\subsection{Fitting and extrapolation procedure}
\begin{figure}[o]
	\centering
		\includegraphics{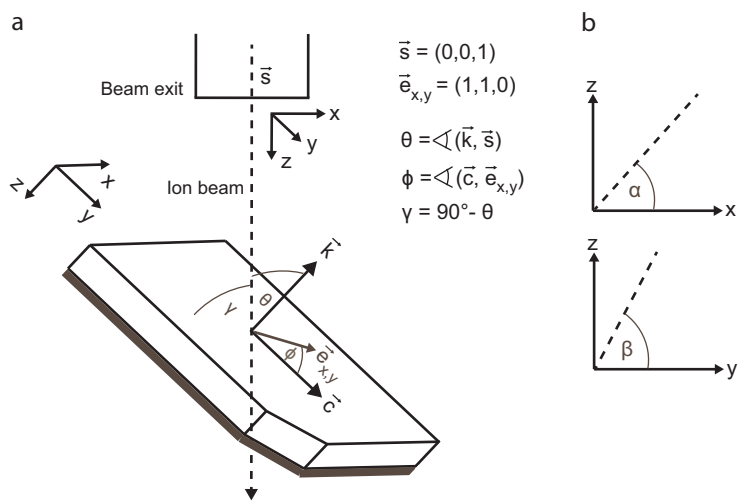}
	\caption{Irradiation setup and particle track reconstruction. (a) The ions are traversing the detector under the angle $\theta$ - the angle between direction of propagation of the ions $\vec{s}$ and k-axis of the FNTD (of dimension 4x8x0.5 $\mu m^{3}$). $\vec{k},\vec{c},\vec{e}_{x,y}$ and all angles refer to the coordinate system of the FNTD. $\vec{s}$ refers to the beam coordinate system. The cell-coating is indicated by the gray layer. (b) Splitting of fitting procedure into two separate linear regression analysis for the x- and y-coordinate of the track spot centers.}
	\label{fig:setup} 
\end{figure}

For reconstruction of the ion trajectory in 3D we assumed the ion to follow a straight line. A line was expected as due to the high energies of the ions, multi-coulomb scattering of the projectile in the crystal lattice is very small. Due to statistical variation in energy deposition and inhomogenities of the color-center density, only uncertainties in the horizontal coordinates (x,y) of the ion track centers were taken into account. Uncertainty in z by the high-precision z-piezo stage (in the nm range) were neglected. The fitting procedure was split up into two separate linear regression analysis (LRA, using least-square estimation \cite{Montgomery2006}) for the x- and y-coordinates respectively (figure \ref{fig:setup}b): 
\begin{eqnarray} 
 x(z) &=& a_{x} + b_{x} \cdot z \label{eq:x_z} \\
 y(z) &=& a_{y} + b_{y} \cdot z \label{eq:y_z}
\end{eqnarray}
The flight direction of the incident ions was parameterized by $\theta$ and $\phi$. $\theta$ (figure \ref{fig:setup}a) was calculated by 
\begin{eqnarray}
\theta &=& 90 \degree - \gamma, \\
\gamma &=& \arctan[(\tan^{-2}\alpha + \tan^{-2}\beta)^{-0.5}]
\end{eqnarray}
with $\tan\alpha = b_{x}$ and $\tan\beta= b_{y}$. $\phi$ was calculated by 
\begin{equation}
\phi = \arctan[ \tan(\alpha) / \tan(\beta)]. 
\end{equation}
Fitting of the particle tracks was carried out in a total range in z of 60 $\mu$m partly starting at different depths in the FNTD. This is the range allowing to track a traversing ion within a single imaging field (135x135 $\mathrm{\mu m^2}$) under angular irradiation (for greater ranges the imaging field has to be moved, i.e. tile scans have to be performed).\\
The ion trajectories obtained by fitting were extrapolated into the A549 cell-layer of 10 $\mu m$ thickness (nominal thickness without correction for distortion, see section \textit{Correction for axial geometrical distortion}, below) grown on top of the FNTD. The 95$\%$ prediction intervals $\mathrm{PI_{x,y}}$ on future observation $(x_{0},y_{0})$ with $\mathrm{PI_{x,y}}= [x_{0} - a_{x} - b_{x} \cdot z_{0}, y_{0}- a_{y} - b_{y} \cdot z_{0}]$ were calculated by
\begin{eqnarray} 
PI_{x} &=& s \cdot t_{n-2,97.5} \cdot \sqrt{1 + \frac{1}{n} + \frac{1}{\hat{s}}(z_{0}-\bar{z})^{2}} \label{eq:PIx} \\
s &=& \sqrt{\frac{\sum_{i=1}^{n}{(a_{x} + b_{x} \cdot z_{i}  - x_{i})^{2}}}{n-2}}	\label{eq:PI}\\
\hat{s} &=& \sum_{i=1}^{n}{(z_{i} - \bar{z})^2}	\\
\bar{z} &=& \frac{1}{n} \cdot \sum_{i=1}^{n}{z_{i}}
\end{eqnarray}
and analog for $PI_{y}$ using $y_i, a_y$ and $b_y$ in (\ref{eq:PI}). Parameter $n$ is the number of track spot centers (with coordinates $x_{i}, y_{i}, z_{i}$) considered for the fit, and $t_{n-2,97.5}$ is the 97.5$\%$ quantile of the t-distribution with $\mathrm{n-2}$ degrees of freedom. We assumed that all extrapolations start at the same depth, 3 $\mu$m below the FNTD crystal surface.\\
Accuracy of ion track reconstruction in 3D was expressed by $\Delta\theta$, $\Delta\phi$ (95$\%$ confidence intervals gained by the LRA) and $\mathrm{PI_{x,y}}$. To study the impact of the parameter $n$ on accuracy we varied the distance in z between two consecutive track spots $\Delta z$ (nominal distance without correction for distortion, see section \textit{Correction for axial geometrical distortion}, below). We further compared three different approaches for the identification of the track spot center coordinates (\itshape iw centroid, rel thres, abs max\upshape, see section \itshape Detection of the center of a track spot\upshape, above):
\begin{enumerate}
	\item{$\Delta$z= 3 $\mu$m, n= 21 track spots}
	\item{$\Delta$z= 6 $\mu$m, n= 11 track spots}
	\item{$\Delta$z= 15 $\mu$m, n= 5 track spots}
\end{enumerate} 
We analysed 20 tracks each for angular and perpendicular irradiated FNTDs.

\subsection*{Correction for axial geometrical distortion}
In the optical path of the confocal imaging a mismatch in index of refraction occurs at the interface between the immersion oil and cell layer and at the interface between cell layer and FNTD. The refractive index of the glass bottom dish and the immersion oil was assumed to equal. The mismatch causes axial distortion of the nominal focal position (position in z without spherical aberration) from the actual one (presence of spherical aberrations) \cite{Jacobsen1995, Elburg2007}. For correction of the negative mismatch ($\mathrm{n_{oil}} = 1.51 > \mathrm{n_{cell}}= 1.47 $ \cite{Hell1993}) occurring at the oil-cell layer boundary we used an improved linear correction method for high NA lenses as previously described in \cite{Elburg2007}. The axial scaling factor (ASF) to recalculate the nominal focal position $z_1$ in the cell layer was determined by:
\begin{equation}
ASF_1= c_1 \Delta n/n_{oil} + c_2 \arctan(c_3 \Delta n/n_{oil}) 
\end{equation}
with $c_1= 1.132$, $c_2= 0.0065$, $c_3= 100$ and $\Delta n= (n_{cell} - n_{oil})$. To account for the positive mismatch at the cell layer-FNTD boundary ($\mathrm{n_{Al_2O_3: C,Mg}} = 1.76 > \mathrm{n_{cell}}$) we used the paraxial approximation $ASF_2 = n_{Al_2O_3: C,Mg}/ n_{cell}$. We used the this approximation as $n_{Al_2O_3: C,Mg}$ exceeds the range of mismatch being considered for the improved linear correction method \cite{Elburg2007} and as it was previously approved to approximate experimental results well \cite{Hell1993}. The position in z of the actual focal point $z_{act}$ (origin is at the bottom of the culture dish) was then calculated by
\begin{equation}
z_{act}(z)= ASF_1 \cdot z_1 + ASF_2 \cdot z 
\end{equation}
with $z_1$ being the nominal thickness of the cell layer (distance between the dish bottom and the detector surface) and $z$ being the nominal position in the FNTD (measured from the detector surface). The aqueous medium between dish bottom and cells was neglected in the correction for axial distortion.\\
Uncertainties in the position of the surface $\Delta s$ translate into an error $\hat{x}$ of the location of the track in the horizontal:
\begin{eqnarray}
z_{act}(z, \Delta s) &=& z_{act}(z) \pm (ASF_1 - ASF_2)\Delta s	 \label{eq:z_act} \\
\hat{x} &=& (ASF_1 - ASF_2)\Delta s / \tan(90\degree - \theta) \label{eq:x_hat}
\end{eqnarray}
starting from
\begin{equation}
z_{act}(z)= ASF_1(s-o_1) + ASF_2 \left[ (\hat{n}-1)\widehat{\Delta z} - (s-o_2)\right]
\end{equation}
and defining $z:= (\hat{n}-1)\Delta z$ as a function of the actual plane number $\hat{n}$ of the acquired image stack and the interval between two consecutive image planes $\widehat{\Delta z}$. $o_1$ and $o_2$ are the respective positions of the first image plane of the cell and FNTD image stack. 

\section{Results}

\subsection{Intensity profile of a track spot}
\begin{figure}[u]
	\centering
		\includegraphics{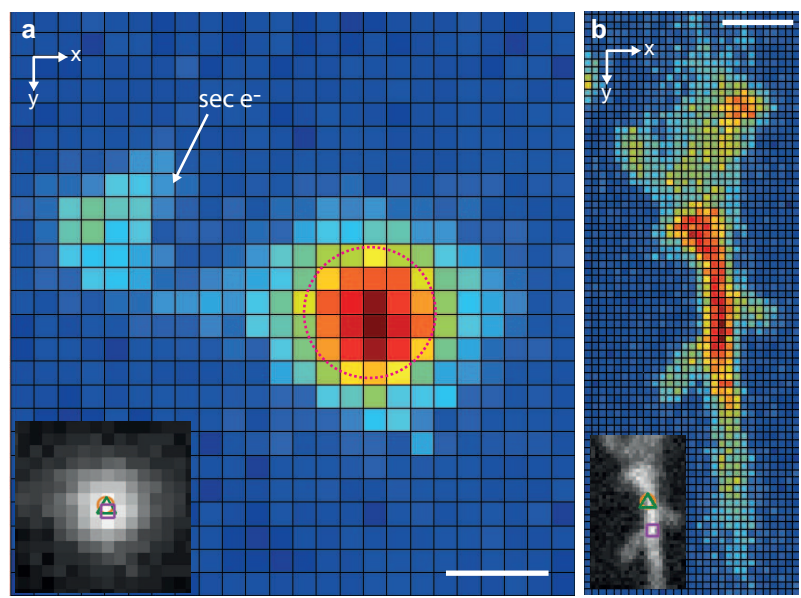}
		\caption{Intensity profile of FNTD read-out signal after (a) perpendicular ($\theta= 0\degree$) and (b) angular irradiation ($\theta= 60\degree$). The track core (masked by the point spread function of the imaging system \cite{NiklasLumDetr2012}) is indicated by the red dashed circle. In both cases secondary electron trajectories are sprouting from the symmetrical and ellipsoidal track spot. Inserts: The track spot centers detected by identifying the intensity-weighted centroid applying a manually defined threshold (ocher circle), the absolute maximum (magenta rectangle) and by identifying the intensity-weighted centroid applying dynamic thresholding (green triangle) do not coincide. Scale bars, (a) 0.5 $\mu$m  and (b) 1 $\mu$m.}
	\label{fig:profile} 
\end{figure}

In figure \ref{fig:profile} track spots - the ion's footprint left in the $\aluminaM$ crystal - resulting from irradiation perpendicular ($\theta = 0 \degree$) and angular ($\theta = 60 \degree$) towards the FNTD surface are shown. Under perpendicular irradiation the track spots have an almost radial symmetric intensity profile with a steep intensity gradient in the track core (halo around the track spot center) and branching trajectories from secondary electrons of higher energy. Their local intensity maxima are due to the track ends.\\
Under angular irradiation the track spots are deformed into ellipsoidal objects extended along beam direction (x) with tattered edges and with branching secondary electron trajectories. The electron trajectories exhibit a broad angular distribution causing a more complex geometry of the track spots. The track spots also exhibit an intensity rise with a global maximum most probably accounting for the angular track core. On average an elongated track spot has the size of approximately 7 x 0.5 $\mathrm{\mu m^2}$ - compared to the diameter of 0.9 $\mu$m of a symmetrical track spot.\\
The inserts in figure \ref{fig:profile}a,b show the examples of the track spot centers for all three identification approaches (\itshape iw centroid, rel thres, abs max, \upshape see section \textit{Fitting and extrapolation procedure}, above). Under perpendicular and angular irradiations the positions of the intensity-weighted track spot centers detected by applying a global threshold and dynamic thresholding are nearly identical. Larger deviations from these positions occur for identifying the absolute intensity maximum. The spatial mismatch is much more pronounced under angular than under perpendicular irradiation.   

\subsection{Residuals of the fit}
\begin{figure}[o]
	\centering
		\includegraphics{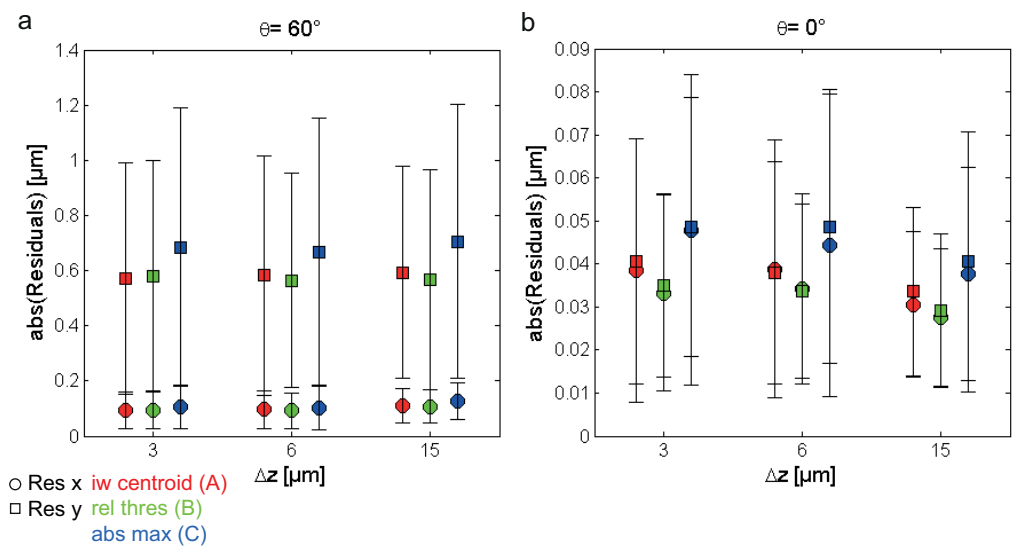}
		\caption{Mean absolute residuals in x (circles) and y (solids) resulting from different fitting procedures. The distance $\Delta z$ between two consecutive track spots was varied and different approaches to detect the track spot centers were used (red: intensity-weighted centroid, green: dynamic thresholding, blue: absolute maximum). The error bars are the standard errors of the mean. (a) angular irradiation, $\theta= 60\degree$ (b) perpendicular irradiation, $\theta= 0\degree$.}
	\label{fig:res} 
\end{figure}

For all LRA (angular and perpendicular irradiation) the residual plots generally displayed a normal distribution (not shown). Figure \ref{fig:res} presents the mean absolute residuals (in an initial step the absolute residuals corresponding to a single fit were averaged) including the standard error of the mean (SEM).\\
In the case of angular irradiation the residuals in x (perpendicular to beam direction) and in y (along beam direction) resulting from identification approach A and B (\itshape iw centroid, rel thres\upshape) fluctuate around 0.10 $\mu$m and 0.58 $\mu$m respectively for all $\Delta z$ (figure \ref{fig:res}a). The residuals concerning identification of the absolute maximum (\itshape abs max\upshape, approach C) are larger (x: $\approx 0.12$ $\mu$m, y: $\approx$ 0.68 $\mu$m). SEMs in x ($\leq 0.02$ $\mu$m) and y ($\approx$ 0.10 $\mu$m) approximately equal for all $\Delta z$ and all identification approaches; the largest values are obtained for approach C.\\
For perpendicular irradiation the residuals in x and in y, resulting from identification approach A and B ($\Delta z= 3$ and 6 $\mu$m) fluctuate between 0.03 $\mu$m and 0.04 $\mu$m and decrease for $\Delta z$= 15 $\mu$m (figure \ref{fig:res}b). Concerning identification approach C the residuals in x and in y lie between 0.04 $\mu$m and 0.05 $\mu$m. All SEMs are smaller than 0.01 $\mu$m.

\subsection{Accuracy of ion track reconstruction in 3D}
\begin{figure}[u]
	\centering
		\includegraphics{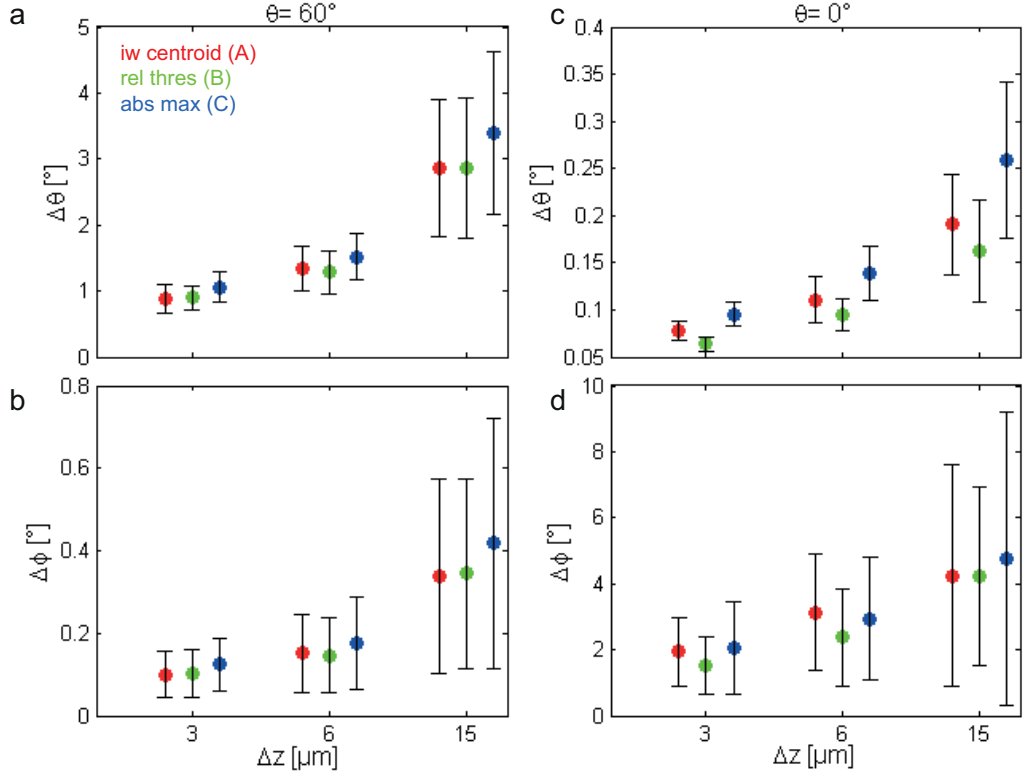}
		\caption{Errors $\Delta\theta$ and $\Delta\phi$ of the different fitting procedures (red: intensity-weighted centroid, green: dynamic thresholding, blue: absolute maximum). $\Delta z$ is the distance in z between two track spots. (a)-(b) angular irradiation, $\theta= 60\degree$, (c)-(d) perpendicular irradiation, $\theta= 0\degree$. The error bars are the standard deviations.}
	\label{fig:angle}  
\end{figure}

\begin{figure}[o]
	\centering
		\includegraphics{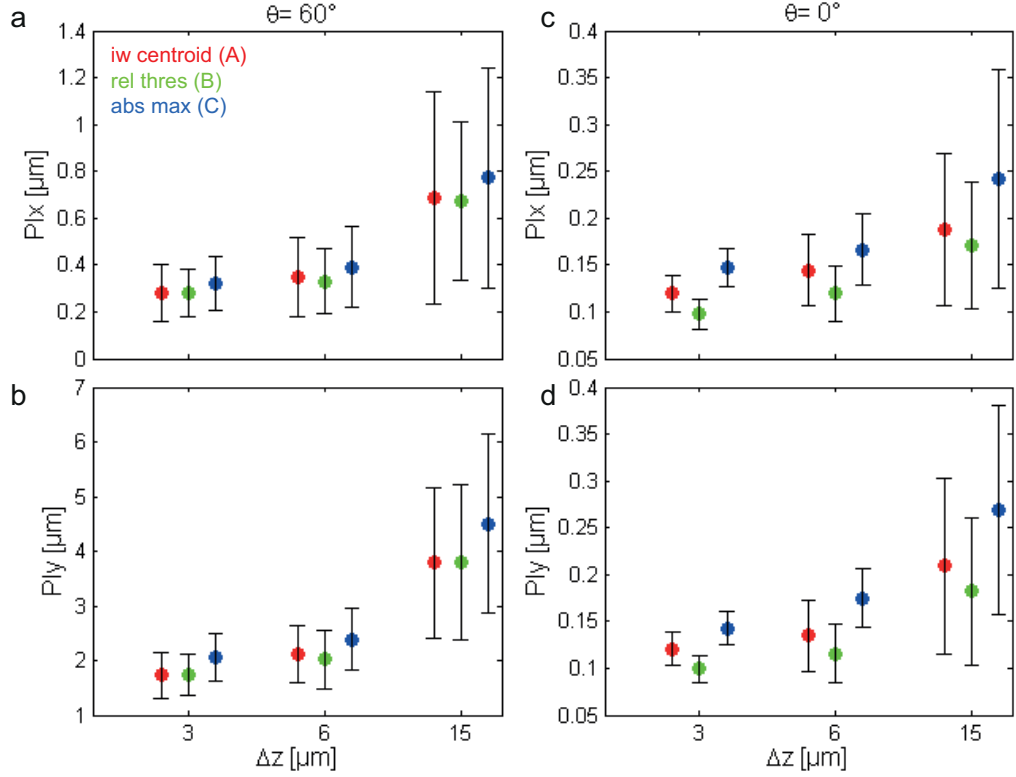}
		\caption{Mean prediction intervals (PI) for x and y resulting from the extrapolation of the particle track into a cell layer of 10 $\mu$m thickness for different fitting procedures (red: intensity-weighted centroid, green: dynamic thresholding, blue: absolute maximum). $\Delta z$ is the distance in z between two track spots. (a)-(b) angular irradiation, $\theta= 60\degree$, (c)-(d) perpendicular irradiation, $\theta= 0\degree$. The error bars are the standard deviations.}
	\label{fig:PI}
\end{figure}

For angular irradiation the mean values $\theta= 59.03\degree$ and $\phi= 84.68\degree$ defining the direction of the traversing ions differ of maximum 0.07$\degree$ and 0.01$\degree$ respectively for all LRA (table 1). The error $\Delta\theta$, figure \ref{fig:angle}a, is smallest (0.89$\degree$) for approach A at $\Delta z=$ 3 $\mu$m and is increasing to 3.45$\degree$ for approach C at $\Delta z=$ 15 $\mu$m. Except for $\Delta z$= 15 $\mu$m, $\Delta\phi$ is smaller than 0.2$\degree$ for all LRA (figure \ref{fig:angle}b). Concerning the mean PIs (figure \ref{fig:PI}a,b and table 1) in the cell-layer of 10 $\mu$m thickness on top of the FNTD the values in x are smaller than 0.4 $\mu$m ($\Delta z < 15$ $\mu$m, for all identification approaches) and are increasing to 0.77 $\mu$m for approach C at $\Delta z$= 15 $\mu$m. $\mathrm{PI_y}$ is has its lowest value (1.72 $\mu$m) for approach A at $\Delta z=$ 3 $\mu$m and is increasing to 4.58 $\mu$m for approach C at $\Delta z=$ 15 $\mu$m.\\
Concerning perpendicular irradiation, $\theta \approx 1.22\degree$ differs less than 0.02$\degree$ for all LRA (table 2). Azimuthal angle $\phi$ lies between 45$\degree$ and 46$\degree$. For $\Delta z <$ 15 $\mu$m and all identification approaches $\Delta\theta$ is smaller than 0.15$\degree$ and is increasing to 0.26$\degree$ for $\Delta z =$ 15 $\mu$m (approach C, figure \ref{fig:angle}c). $\Delta\phi$ covers a broad range - between 1.53$\degree$ and 4.77$\degree$ (figure \ref{fig:angle}d). The mean PIs in x and in y are generally much smaller than 0.3 $\mu$m (figure \ref{fig:PI}c,d and table 2).\\
Both, the distributions of $\Delta\theta$ and $\Delta\phi$ (perpendicular and angular irradiation) indicated by their standard deviation (s.d.) and mean values show a steep rise between $\Delta z =$ 6 $\mu$m and $\Delta z =$ 15 $\mu$m. This is independent of the identification approach used (figure \ref{fig:angle}). The effect is especially pronounced under angular irradiation. The mean PIs behave similarly (figure \ref{fig:PI}).

\section{Discussion}
Accuracy of ion track reconstruction in 3D using the FNTD depth information is strongly influenced by the irradiation angle $\theta$ and therefore by complexity of the intensity profile of a track spot as well as by the approach for identification of its center.

\subsection{Intensity profile of a track spot}
The size and shape of a track spot depends on the energy of the incident ions \cite{Akselrod2011, NiklasLumDetr2012} and $\theta$ governing the geometrical cross section of the traversing particles with the detector material. High-energy carbon ion irradiation produces a high density of secondary electrons of low energies responsible for the generally symmetrical as well as ellipsoidal intensity profile of the track spots (figure \ref{fig:profile}). The sprouting electron trajectories with a broad angular distribution are caused by $\delta$ electrons having enough energy to leave the track-core. These trajectories, although less probable, and their random formation (due to frequent scattering of the $\delta$ electrons) distort the original symmetrical intensity profile.\\
The track spots under angular irradiation seem to comprise more sprouting electron trajectories than the symmetrical track spots. Due to a greater geometrical cross section the probability of producing fast $\delta$ electrons increases including the formation of tattered track spot edges. In addition, electron trajectories of neighboring ion traversals from above or below crossing as well as being scattered in the image plane are being detected. On the contrary, under perpendicular irradiation mainly the electron trajectories perpendicular to the flight direction of the incident ion are visible.\\ 
The track spot is masked by the point spread function (PSF) of the imaging system (figure \ref{fig:profile}) \cite{NiklasLumDetr2012}. The PSF is distorted by the refractive index mismatch at the oil-cell layer and cell layer-$\aluminaM$ interface causing spherical aberrations \cite{Carlsson1991, Hell1993, Jacobsen1995}. Besides a decrease of image brightness (a result of the confocal arrangement), the aberrations could also account for a blurred appearance of the track spots (including the tattered edges) by an increasing FWHM of the PSF and appearance of secondary maxima \cite{Hell1993}. A deconvolution could increase resolution only to some extend as the PSF in the corresponding focal depth and the actual structure of a track spot have to be exactly known.

\subsection{Ion track center assessment}
The thresholds based methods A and B (\itshape iw centroid, rel thres\upshape) using a value of 0.4 for identifying intensity-weighted centroids and 2/3 of the maximum pixel value within an ROI (dynamic thresholding) showed best suitability for carbon ion irradiation and minimize the probability to detect $\delta$ electron structures (including their Bragg-peaks) as well as the tattered track spot edges. As the gradient of the intensity profile is relatively steep, 2/3 of the maximum is not too high running the risk of approaching the global maximum which is likely not to coincide with the center of the track spot core. Statistical fluctuation of energy deposition and local fluctuation of the color center density could be the main reason for this mismatch. Weighting by intensity seems to approach the actual track core best. This coincides well with smaller residuals (figure \ref{fig:res}) and hence more accurate track reconstruction (figures \ref{fig:angle} and \ref{fig:PI}) than the detection of the absolute intensity maximum (\itshape abs max\upshape, method C). In addition, dynamic thresholding has the advantage of responding to different intensity profiles (variation in the maximum intensity \cite{NiklasLumDetr2012}) by different particle types or energies within a FNTD image.\\
Under angular irradiation it is more difficult to define the actual track spot core. A fitting of the intensity profile of a track spot by a 2D Gaussian function \cite{Akselrod2006, NiklasLumDetr2012} (symmetrical for perpendicular irradiation, elliptical for angular irradiation) and the detection of its maximum could improve the subsequent LRA.

\subsection{Accuracy on particle track reconstruction in 3D}
Complexity of the track spot geometry and hence the irradiation angle $\theta$ have a major impact on accuracy expressed by $\Delta\theta$, $\Delta\phi$ and PI. The radial symmetrical intensity profile ($\theta = 0\degree$) is reflected in nearly identical residuals in x and in y (figure \ref{fig:res}b) for all LRA. On the contrary the elongation (along beam direction) and much higher complexity of the track spots under angular irradiation ($\theta = 60\degree$) causes much greater discrepancy between the residuals in x and in y - by a factor of approximately six - irrespective of the identification approach of the track spot center (figure \ref{fig:res}a). This is directly reflected in the increased error $\Delta\theta$ by a factor of approximately ten (comparing $\theta = 60\degree$ and $\theta = 0\degree$, figure \ref{fig:angle}a,c). This is also reflected in much greater PIs (figure \ref{fig:PI}), especially for $\mathrm{PI_y}$ under angular irradiation being more than ten times larger than the symmetrical PIs under perpendicular irradiation.\\ 
The surprisingly great values of $\Delta\phi$ under perpendicular irradiation (figure \ref{fig:angle}d) arise from the narrow distribution of the projections of the track spots onto the yz plane. Small fluctuations have a great impact leading to large variations in the coefficient interval of $b_{y}$ in the LRA (equation \ref{eq:y_z}). To gain more reliable values and to decrease $\Delta\phi$ below 1$\degree$ we increased the number of track spots by extending the total range in z for the LRA for perpendicular irradiation to 114 $\mu$m with $\Delta$z= 3 $\mu$m (table 3).\\
To principally minimize the PIs and irrespective of the irradiation angle it is favorable to start the FNTD read-out near the detector surface. This assures a minimal extrapolation interval into the cell layer. This also reduces the PIs as the interval between the mean z-coordinate of identified track spots and z-coordinate of extrapolation gets minimized (equation \ref{eq:PIx}).\\
The strong decrease in accuracy (and larger s.d.) for $\Delta z$= 15 $\mu$m, all identification approaches (figure \ref{fig:angle}, \ref{fig:PI}), could arise from the insufficient averaging and the statistical nature of energy deposition. Under perpendicular irradiation a reasonable trade-off between fast detector read-out and high accuracy is yet possible by increasing $\Delta$z to 15 $\mu$m (figure \ref{fig:angle}c). $\Delta\theta$ increases but is still less than 0.3$\degree$. $\Delta\phi$ has little significance. The PIs are however increasing as they depend on the inverse number of track spots considered for the LRA. The surprising decrease of the residuals ($\Delta z$= 15 $\mu$m, all three identification approaches, figure \ref{fig:res}b) has no impact on the increase of $\Delta\theta$, $\Delta\phi$ and PIs.

\subsection{Correction of axial geometrical distortion}
The distortion induced by refractive index mismatch only affects the axial position of the focal point and hence $\theta$ - the trajectory of the particles. Introducing non-linear correction \cite{Elburg2007} could further improve reconstruction accuracy to some extent especially for larger angles. As the refractive index differs for individual organelles \cite{Beuthan1996} the actual correction term for the cell layer has a spatial dependence. To reduce the impact of refractive index mismatching on track reconstruction especially for angular irradiation, 2,2'-Thiodiethanol (98\%, n= 1.518) can be used as mounting medium for the cell layer during the confocal read-out \cite{Staudt2007}. As its refractive index equals with the refractive index of the immersion oil only the oil-$\aluminaM$ boundary has to be considered for axial correction. To further reduce spherical aberrations, perpendicular irradiation should be employed in particular for live cell imaging. Despite the presence of the refractive index mismatch at the cell-$\aluminaM$ boundary the nominal can be approximated with the actual focal position.\\
Concerning uncertainty in the position of the FNTD surface $\Delta s$, the error term $(ASF_1 - ASF_2)\Delta s$ in (\ref{eq:z_act}), which affects the actual focal position $z_{act}$ is independent of $z$ and does not influence the LRA and thus trajectory of the incident ions (parameterized by $\theta$ and $\phi$). The resulting error $\hat{x}$ (\ref{eq:x_hat}) of the position of the ion track in the horizontal direction is governed by $\theta$ and the product $(ASF_1 - ASF_2)\Delta s$. $\Delta s = 1 \mu m$ at $\theta= 60 \degree$ yields $\hat{x}= 0.40$ $\mu m$. For shallow irradiation angles, i.e. $\theta < 45\degree$, $\hat{x}$ decreases steeply as long as $(ASF_1 - ASF_2) < 1$. $\hat{x}$ becomes yet more relevant if the ion traversal in the intracellular space has to be localized.\\

\section{Conclusions}
It is possible to perform accurate ion track reconstruction in 3D and extrapolate an ion's trajectory into a cell layer covering the FNTD by using depth information provided by detector read-out. The accuracy of the track reconstruction procedure strongly depends on the irradiation angle $\theta$ and by the approach used to identify individual track spot centers. The use of intensity-weighted centroids and dynamic thresholding yield the highest accuracy. Steep irradiation angles distort the otherwise symmetrical track spots, resulting in reduced accuracy of centroid prediction. To achieve the desired accuracy in determining the angle $\theta$ with an error smaller than 1$\degree$ under angular irradiation at $\theta= 60\degree$ the intensity-weighted centroid detection and a 21 image stack separated by $\widehat{\Delta z}= 3$ $\mu$m was found as the best set of image acquisition and processing parameters. Increasing $\widehat{\Delta z}$ to 6 $\mu$m (i.e. half the number of track spots) deteriorates the accuracy, i.e. $\Delta\theta$ and PI, roughly by 30$\%$ and 20$\%$ respectively. Concerning angles of irradiation with $\theta$ close to 0$\degree$ and using the same fit parameters, $\Delta\theta$ decreases by a factor of at least 10 with nearly symmetrical lateral error distribution.\\
The required increase in accuracy of track reconstruction also increases  the total read-out time of the detector. To further improve accuracy if necessary one could extend the total axial range of the image stack, to increase the number of images and decrease the image depth increment between two consecutive image planes ($\widehat{\Delta z}$). This, however, is more time consuming. Naturally the total microscope time is directly affected by the acquisition time of a single image plane (mainly governed by laser power $p$, dwell time $\tau$, number of rescans $R$ and total number of spots per imaging field \cite{Greilich2012}) influencing the signal-to-noise ratio and thus the quality of the detection of the track spot center.\\
The ability of FNTD technology to provide accurate 3D ion track reconstruction makes it a powerful tool for radiobiological investigations in clinical ion beams, either being used as a substrate to be coated with living tissue (cell-fluorescent ion track hybrid detector) \cite{NiklasRadOnc2012} or being implanted in vivo.

\section*{Author's contributions}
MN performed experiments, analyzed data and wrote the manuscript.  JB participated in the axial geometrical correction. MA developed the crystal (FNTD) material and FNTD imaging technique. MN, AA, OJ and SG designed experiments and interpreted the data. All authors edited the paper. All authors read and approved the final manuscript.

\section*{Acknowledgements}
We are grateful to S. Brons for generously providing support and technical irradiation assistance at the Ion Beam Therapy Center of Heidelberg University Hospital. We should also like to thank F. Bestvater and M. Brom of the DKFZ's light microscopy core facility  for their enthusiasm and unflagging support. MN is funded by the fellowship from the Helmholtz International Graduate School for Cancer Research at the German Cancer Research Center. SG and AA are supported by the German Cancer Researc Center. OJ is supported by the University Hospital Heidelberg. MA is supported by Landauer R$\&$D money. J.B. is supported by Oklahoma State University. This work was supported by the Helmholtz Association (Translating hadron therapy from basic research to clinical application, VH-VI-303, S.G.), German Research Council (DFG,  KFO214), National Aeronautics and Space Administration under NSCOR grant no. NNJ06HA28G, the German Krebshilfe (Deutsche Krebshilfe, Max-Eder 108876), intramural Grants of the National Center for Tumor diseases (NCT, Heidelberg, Germany) and the German Federal Ministry  of Research and Technology (Bundesministerium f\"{u}r Bildung und Forschung - BMBF 03NUK004C). The authors declare that they have no competing financial interests.



\begin{table}[u]
\caption{\label{angularIrr}Angular irradiation, $\theta= 60\degree$. Accuracy of ion track reconstruction using different fitting procedures. $\Delta z$: distance in z between two track spots. The errors are the s.d. Total range in z: 60 $\mu$m.}
\begin{indented}
\item[]\begin{tabular}{@{}lllllll}
\br
& $\theta [\degree]$ & $\Delta\theta [\degree]$ & $\phi [\degree]$ & $\Delta\phi [\degree]$ & $\mathrm{PI_x}$ [$\mu$m] & $\mathrm{PI_y}$ [$\mu$m]\\
\br
$n= 21$ track spots,&  $\Delta z = 3 \mu m$ \\
\mr
      iw centroid (A)& $58.97$ & $0.89 \pm 0.22$ & $ 84.69 $ & $0.10 \pm 0.06$ & $0.28 \pm 0.12$ & $1.72 \pm 0.43$  \\ 
      rel thres (B)& $58.98$ & $0.90 \pm 0.19$ & $84.69$ & $0.10 \pm 0.06$ & $0.28 \pm 0.10$ & $1.74 \pm 0.37$   \\ 
      abs max (C)& $58.98$ & $1.06 \pm 0.23$ & $84.69$ & $0.13 \pm 0.06$ & $0.32 \pm 0.11$ & $2.07 \pm 0.44 $  \\ 
\br
$n= 11$ track spots,&  $\Delta z = 6 \mu m$ \\
\mr
    	iw centroid (A) & $59.00$ & $1.36 \pm 0.34$ & $84.68$ & $0.16 \pm 0.09$ & $0.35 \pm 0.17$ & $2.14 \pm 0.54$  \\ 
    	rel thres (B)& $59.01$ & $1.23 \pm 0.33$ & $84.68$ & $0.15 \pm 0.09$ & $0.33 \pm 0.13$ & $2.01 \pm 0.53$   \\     	
    	abs max (C)& $59.03$ & $1.54 \pm 0.37$ & $84.68$ & $0.18 \pm 0.11$ & $0.38 \pm 0.17$ & $2.42 \pm 0.58 $  \\ 
\br
$n= 5$ track spots,&  $\Delta z = 15 \mu m$ \\
\mr
    	iw centroid (A)& $59.08$ & $2.84 \pm 1.03$ & $84.67$ & $0.34 \pm 0.23$ & $0.68 \pm 0.45$ & $3.76 \pm 1.37$  \\ 
    	rel thres (B)& $59.09$ & $2.79 \pm 1.08$ & $84.68$ & $0.33 \pm 0.24$ & $0.66 \pm 0.33$ & $3.69 \pm 1.45$   \\    	
    	abs max (C)& $59.09$ & $3.45 \pm 1.21$ & $84.68$ & $0.43 \pm 0.29$ & $0.77 \pm 0.47$ & $4.58 \pm 1.62 $  \\ 
\br
\end{tabular}
\end{indented}
\end{table}

\begin{table}[u]
\caption{\label{perpendIrr}Perpendicular irradiation, $\theta= 0\degree$. Accuracy of ion track reconstruction using different fitting procedures. $\Delta z$: distance in z between two track spots. The errors are the s.d. Total range in z: 60 $\mu$m.}
\begin{indented}
\item[]\begin{tabular}{@{}lllllll}
\br
& $\theta [\degree]$ & $\Delta\theta [\degree]$ & $\phi [\degree]$ & $\Delta\phi [\degree]$ & $\mathrm{PI_x}$ [$\mu$m] & $\mathrm{PI_y}$ [$\mu$m]\\
\mr
$n= 21$ track spots,&  $\Delta z = 3 \mu m$ \\
\mr
	  iw centroid (A) & $1.20$ & $0.08 \pm 0.01$ & $45.16$ & $1.95 \pm 1.04$ & $0.12 \pm 0.02$ & $0.12 \pm 0.02$ \\ 
	  rel thres (B) & $1.21$ & $0.06 \pm 0.01$ & $44.99$ & $1.53 \pm 0.87$ & $0.10 \pm 0.02$ & $0.10 \pm 0.01$ \\ 	  
	  abs max (C) & $1.22$ & $0.10 \pm 0.01$ & $45.30$ & $2.05 \pm 1.42$ & $0.15 \pm 0.02$ & $0.14 \pm0.02$  \\ 
\br
$n= 11$ track spots,& $\Delta z = 6 \mu m$ \\
\mr
	  iw centroid (A)& $1.21$ & $0.11 \pm 0.02$ & $44.95$ & $3.14 \pm 1.78$ & $0.14 \pm 0.04$ & $0.13 \pm 0.04$ \\ 
	  rel thres (B)& $1.22$ & $0.09 \pm 0.02$ & $44.92$ & $2.38 \pm 1.47$ & $0.12 \pm 0.03$ & $0.12 \pm 0.03$ \\ 	  
	  abs max (C)& $1.24$ & $0.14 \pm 0.03$ & $45.94$ & $2.94 \pm 1.86$ & $0.17 \pm 0.04$ & $0.17 \pm0.03$  \\ 	  
\br
$n= 5$ track spots,&  $\Delta z = 15 \mu m$ \\
\mr
	  iw centroid (A)& $1.21$ & $0.19 \pm 0.05$ & $45.21$ & $4.23 \pm 3.36$ & $0.19 \pm 0.08$ & $0.21 \pm 0.09$ \\ 
	  rel thres (B)& $1.22$ & $0.16 \pm 0.05$ & $45.39$ & $4.21 \pm 2.71$ & $0.17 \pm 0.07$ & $0.18 \pm 0.08$ \\ 	  
	  abs max (C)& $1.24$ & $0.26 \pm 0.08$ & $45.70$ & $4.77 \pm 4.45$ & $0.24 \pm 0.12$ & $0.27 \pm 0.11$  \\   
\br
\end{tabular}
\end{indented}
\end{table}

\begin{table}[u]
\caption{\label{perpendIrr114}Perpendicular irradiation, $\theta= 0\degree$. Accuracy of ion track reconstruction using different fitting procedures. $\Delta z$: distance in z between two track spots. The errors are the s.d. Total range in z: 114 $\mu$m.}
\begin{indented}
\item[]\begin{tabular}{@{}lllllll}
\br
& $\theta [\degree]$ & $\Delta\theta [\degree]$ & $\phi [\degree]$ & $\Delta\phi [\degree]$ & $\mathrm{PI_x}$ [$\mu$m] & $\mathrm{PI_y}$ [$\mu$m]\\
\mr
$n= 39$ track spots,&  $\Delta z = 3 \mu m$ \\
\mr
	  iw centroid (A)& $1.17$ & $0.03 \pm 0.005$ & $45.63$ & $0.74 \pm 0.45$ & $0.11 \pm 0.02$ & $0.11 \pm 0.02$ \\  
	  rel thres (B)& $1.17$ & $0.03 \pm 0.005$ & $45.53$ & $0.63 \pm 0.42$ & $0.10 \pm 0.02$ & $0.10 \pm 0.01$ \\  
	  abs max (C)& $1.18$ & $0.04 \pm 0.005$ & $45.53$ & $0.84 \pm 0.61$ & $0.14 \pm 0.02$ & $0.14 \pm 0.02$  \\	
\br
\end{tabular}
\end{indented}
\end{table}

\end{document}